\documentclass[twocolumn,times,trackchanges, dvipsnames]{aastex631}
\usepackage[utf8]{inputenc}
\usepackage{amsmath}

\begin{document}
\title{An FUV-detected Accretion Shock at the Star-Disk Boundary of FU Ori}

\author{Adolfo S. Carvalho}
\affiliation{Department of Astronomy; California Institute of Technology; Pasadena, CA 91125, USA}
\author{Lynne A. Hillenbrand}
\affiliation{Department of Astronomy; California Institute of Technology; Pasadena, CA 91125, USA}
\author{Kevin France}
\affiliation{Laboratory for Atmospheric and Space Physics, University of Colorado Boulder, Boulder, CO 80303, USA}
\author{Gregory J. Herczeg}
\affiliation{Kavli Institute for Astronomy and Astrophysics, Peking University, Beijing 100871, People's Republic of China}
\affiliation{Department of Astronomy, Peking University, Beijing 100871, People's Republic of China}

\begin{abstract}
    FU Ori objects are the most extreme eruptive young stars known. Their 4 to 5 magnitude photometric outbursts last for decades and are attributed to a factor of up to 10,000 increase in the stellar accretion rate. The nature of the accretion disk-to-star interface in FU Ori objects has remained a mystery for decades. To date, attempts to directly observe a shock or boundary layer have been thwarted by the apparent lack of emission in excess of the accretion disk photosphere down to $\lambda = 2300$ \AA. We present a new NUV and the first high-sensitivity FUV spectrum of FU Ori. The FUV continuum is detected for the first time and, at $\lambda = 1400$ \AA, is more than $10^4$ times brighter than predicted by a viscous accretion disk. We interpret the excess as arising from a shock at the boundary between the disk and the stellar surface. We model the shock emission as a blackbody and find that the temperature of the shocked material is $T_\mathrm{FUV} \approx 16,000 \pm 2000$ K. The shock temperature corresponds to an accretion flow along the surface of the disk that reaches a velocity of 40 km s$^{-1}$ at the boundary, consistent with predictions from simulations.
\end{abstract}


\section{Introduction}\label{sec:introduction}

FU Ori is a young stellar object (YSO) that underwent an enormous photometric outburst in 1937 \citep{Wachmann_FUOri_1954ZA, Herbig_InterpOfFUOri_1966VA}. As other outbursting YSOs were discovered in the following decades \citep{herbig_eruptive_1977}, the nature of the outbursts was eventually understood as a sudden increase in stellar mass accretion rate \citep{hartmann_fu_1996}. FU Ori became the prototype for these outbursting YSOs, and FU Ori objects came to represent the most rapidly accreting YSOs known. 

In this interpretation, FU Ori objects are classical T Tauri stars (CTTSs) that have undergone a disk instability and, as a result, their accretion rates increased from the typical CTTS rate of $10^{-8} \ M_\odot$ yr$^{-1}$ to $10^{-4} \ M_\odot$ yr$^{-1}$. 
\citet{Kenyon_FUOri_disks_1988ApJ} demonstrated that the absorption line profiles and spectral energy distributions from FU Ori objects were consistent with those predicted by a modified version of the \citet{Shakura_sunyaev_alpha_1973A&A} viscous accretion disk model. 
However, unlike the situation for the \citet{Shakura_sunyaev_alpha_1973A&A} theory that had been developed for black hole 
disk accretion, CTTS have finite radius.  This necessitates both a different form for the temperature profile at small radii in the disk,
and revised expectations regarding a ``boundary layer" between the disk and the star \citep{Lynden-Bell_Pringle_ViscousDisks_1974MNRAS, Pringle_accretionDisksReview_1981ARA&A}.

For CTTS, it is generally accepted that accreting material is transported along stellar magnetic field lines 
and deposited near the poles of the star.  Observed blue optical and ultraviolet excess emission,
relative to non-accreting young stars of the same spectral type, is then understood as 
matter shocking onto the stellar photosphere at free-fall velocities 
\citep{Uchida_MagnetosphericAccretion_1984PASJ, koenigl_DiskAccretionTTSs_1991ApJ, valenti_tts_1993,Calvet_FunnelFlowStructure_1998ApJ}.
For FU Ori objects the detailed structure of the accretion boundary layer, which is expected to be equatorial, remains largely unconstrained. 


\citet{Popham_boundaryLayersInPMSDisks_1993ApJ} developed a boundary layer model in which the angular momentum transport efficiency onto the star governed the excess heating at the disk-star interface and led to different temperature profiles parameterized by that efficiency. They then used their model to predict the resulting temperature profiles and observed SEDs and line profiles of FU Ori objects \citep{Popham_boundaryLayerSpectraLineProfiles_1996ApJ}. A comparison with visible range observations favored temperature profiles resembling that constructed by \citet{Kenyon_FUOri_disks_1988ApJ}, which had $T_\mathrm{max} = 6800$ K. 

Between 1979 and 1987, near- and far-ultraviolet (NUV/FUV) spectra of FU Ori were taken by the International Ultraviolet Explorer (IUE). The FUV spectrum\footnote{We visually inspected the FUV spectrum in the IUE Newly Extracted Spectra (INES) archive at \url{http://ines.oat.ts.astro.it/} without further processing. The published spectrum can be seen in \citet{Valenti_IUE_SurveyOfYSOsI_2000ApJS}} revealed line emission from the \ion{Si}{4}, \ion{C}{4}, and \ion{He}{2} features, which are typically interpreted as emission from the accretion shock in CTTSs \citep{JohnsKrull_IUE_SurveyOfYSOsII_2000ApJ, Ardila_hotLinesInCTTSs_2013ApJS}, although the observations of FU Ori were not sensitive enough to analyze the line profiles or to detect the continuum. The NUV spectrum \citep{Kenyon_IUE_FUOri_1989ApJ} indicated a relatively high maximum disk temperature (9,000 K). 

This high temperature complemented theories of a boundary layer region where the heating is dominated by the enormous shear between the disk material, which is orbiting at near-Keplerian speeds, $v_\mathrm{kep} \sim 200$ km s$^{-1}$, and the slower-rotating central star, $v_\mathrm{rot} < 50$ km s$^{-1}$ \citep{Nofi_rotational_2021ApJ}. 
However, the temperature profiles of \citet{Popham_boundaryLayerSpectraLineProfiles_1996ApJ} that give $T_\mathrm{max} \sim 9000$ K 
predict absorption line profiles that are inconsistent with both the optical and ultraviolet observations.

Subsequently, the IUE NUV-derived $T_\mathrm{max} \sim 9000$ K was re-affirmed with a 2001 NUV spectrum of FU Ori taken with HST/STIS. This spectrum is almost identical to the IUE spectrum and \citet{Kravtsova_FUOriSTIS_2007AstL} found that a $T_\mathrm{max} = 9000$ K model is required to match the absorption in the $2300 - 3200$ \AA\ range. Yet, using the radiative transfer model from \citet{zhu_FUOriInnerDisk_2007ApJ} and a lower $A_V = 1.5$ mag (rather than 2.2 mag), \citet{hartmann_FUOriProfile_2011arXiv1106} demonstrate that the NUV continuum can be produced with a much lower $T_\mathrm{max} = 5840$ K.

We present in this Letter the first high-sensitivity FUV spectrum of FU Ori, with a clear continuum detection down to $\lambda \approx 1150$ \AA. The FUV continuum flux is in excess of that predicted by a viscous disk model by a factor of $> 10^4$. We treat the FUV excess as a blackbody and find a $T_\mathrm{FUV} \sim 16,000$ K and a filling factor of 0.02\% on the surface of the $3.52 \ R_\odot$ star. The small emission region and relatively high temperature indicate that the FUV flux arises from shock-heated material where the surface accretion in the disk meets the stellar photosphere. 

In Section \ref{sec:data}, we describe our HST observations and how we discriminate between continuum and line emission. We show our FUV spectrum for both COS and STIS in Section \ref{sec:FUV}. We then introduce our disk model and describe how we fit the observed FUV excess relative to the disk model in Section \ref{sec:cont_fit}. In Section \ref{sec:discussion} we discuss our interpretation of the continuum spectrum in the context of previous boundary layer models.


\section{Data and Calibration} \label{sec:data}
We obtained spectra of FU Ori using the HST Space Telescope Imaging Spectrograph (STIS) and Cosmic Origins Spectrograph (COS) as part of the Guest Observer (GO) program 17176\footnote{The data can be accessed at the Mikulski Archive for Space Telescopes (MAST) via \dataset[doi: 10.17909/3p42-jw31]{https://doi.org/10.17909/3p42-jw31}.}. 

The STIS observations were taken using the $52^{\prime\prime} \times 2^{\prime\prime}$ arcsec slit in the grating settings: G140L (FUV-MAMA), G230L (NUV-MAMA), and G430L. The three settings cover 1150-5500 \AA\ with a typical $R \equiv \lambda/\Delta\lambda = 600$. 

The COS spectra were obtained using the G130M and G160M gratings and 4 central wavelength ($\lambda_\mathrm{cen}$) settings: 1222, 1309, 1589, and 1623 \AA\ to ensure maximal coverage of the FUV at high resolution. The exposure times were 2350, 2348, 2118, and 1666 s, respectively. We used the Primary Science Aperture, which is a spherical aperture with a 2.5$^{\prime\prime}$ diameter.
We combined the four exposures into a single FUV spectrum spanning 1065-1800 \AA\ with a typical $R \sim 15,000$. 
To increase the signal-to-noise of the COS spectrum, we bin by a factor of 5 using the flux-conserving rebinning code $\mathtt{spectres}$ \citep{Carnall_spectres_2017arXiv}. 

The COS spectrum is shown in Figure \ref{fig:COSSpec}. For the purposes of our FUV excess fit, we carefully identified regions of continuum emission, free of emission line contamination, following the method outlined in \citet{France_UVRadiationFields_2014ApJ}. In order to minimize contamination from line emission, we masked any points within $\pm$500 km s$^{-1}$ of atomic lines and $\pm$100 km s$^{-1}$ of H$_2$ lines. The line list we used includes both features we clearly detect in the spectrum of FU Ori and, to be thorough, known bright features in CTTS spectra (regardless of their strength in FU Ori). There are several features with uncertain identification that we masked, and we leave identifying these features to a future paper. The selected continuum regions are shown in light blue in Figure \ref{fig:COSSpec}. 

We then bin those regions further down to 12 points spanning 1150 \AA\ to 1780 \AA\ to represent the FUV continuum flux 
(as indicated in 
Figure \ref{fig:COSSpec}). We assume the uncertainties are normally distributed and propagate them accordingly, including the 5\% COS absolute flux uncertainty\footnote{\url{https://hst-docs.stsci.edu/cosihb/chapter-5-spectroscopy-with-cos/5-1-the-capabilities-of-cos}}. The effective wavelength of each bin is the flux-weighted mean wavelength. As a result, some bins are assigned effective wavelengths that coincide with bright emission lines (e.g., the bin at 1336 \AA), although the bin does not contain flux from that line. We also bin down the 1800 \AA\ to 2400 \AA\ region of the STIS G230L spectrum to the same wavelength spacing to construct a full spectrum of the FUV continuum. This $1150-2400$ \AA\ continuum spectrum is what we use for the model fit described in Section \ref{sec:cont_fit}.

The different apertures between the STIS (52$^{\prime\prime}\times 2^{\prime\prime}$) and COS ($2.5^{\prime\prime}$) do not seem to produce any significant differences between the spectra. Both apertures capture the major features at $<1^{\prime\prime}$ scale in the FU Ori system \citep[see][for a scattered light image as a reference]{Weber_FUOriScatteredLight_2023MNRAS}. The companion, FU Ori S, is detected in line emission in the STIS G140L spectrum. 
We report this in Carvalho et al. (2024, in prep). The binary is clearly resolved in the spatial direction such that extraction of the spectrum of FU Ori N does not include emission from FU Ori S.

\section{The FUV Spectrum of FU Ori} \label{sec:FUV}

The COS spectrum of FU Ori is the first FUV spectrum sensitive enough to detect continuum emission from the source. The continuum is detected from 1170 \AA\ $< \lambda <$ 1780 \AA\ at a signal-to-noise ratio (SNR) greater than 5. The emission blueward of 1170 \AA\ is extremely low SNR and therefore undetected, and redward of 1800 \AA\ is near the detector edges where the spectrum data quality drops rapidly. 

\begin{figure*}[!htb]
    \centering
    \includegraphics[width = 0.93\linewidth]{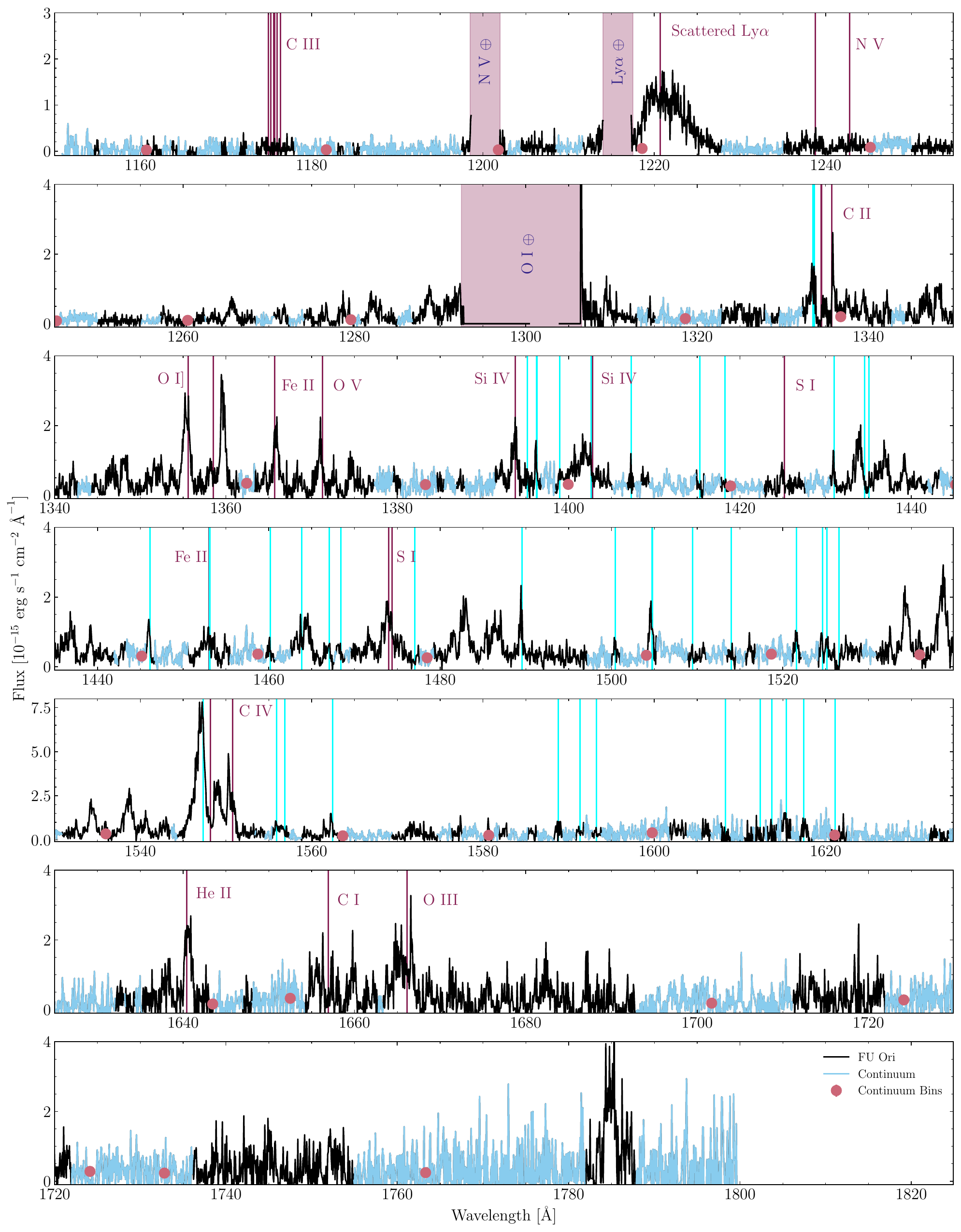}
    \caption{The COS spectrum of FU Ori with the regions selected to represent continuum highlighted in light blue. Regions in black represent those masked out due to observed or potential line emission. The pink points with error bars mark the binned down continuum points we fit in Section \ref{sec:cont_fit}. The vertical cyan lines mark locations of prominent H$_2$ features in typical CTTS spectra \citet{France_RadialDistOfH2_2023AJ}, while identified bright or well-known (but marginally detected) atomic emission lines are marked and labeled in maroon. The vertical shaded regions denote masked geocoronal emission.}
    \label{fig:COSSpec}
\end{figure*}

As can be seen in Figure \ref{fig:COSSpec} the spectrum is line rich, though there are several emission lines that are often seen in CTTS systems but are very weak or only marginally detected here (e.g., \ion{C}{3} 1175 \AA\ multiplet and the \ion{N}{5} 1238.8/1242.8 \AA\ doublet). The detected emission lines typically targeted in FUV studies of CTTSs are marked, as are bright H$_2$ features. 

A more detailed discussion of the line profiles and plots of the individual lines in velocity space are given in Appendix \ref{sec:lineProfs}. As this Letter is focused on the properties of the FUV continuum, we reserve an analysis of the emission line profiles for a future paper.

\begin{figure*}[!htb]
    \centering
    \includegraphics[width = 0.98\linewidth]{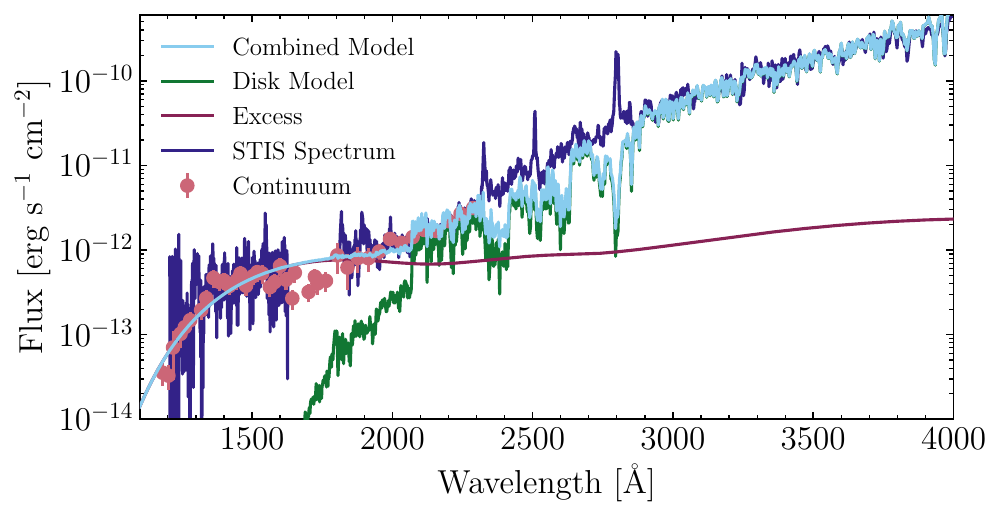}
    \caption{The FUV and NUV STIS spectra (dark blue) and the continuum points we derived from the COS spectrum (pink circles with error bars).  Our combined disk + shock model fit is shown (light blue) along with the individual model components, with the disk model in green and the 16,275 K blackbody component in maroon. Notice the sharp transition from the disk-dominated spectrum to the excess emission dominated spectrum at the 2060 \AA\ continuum break.  The model components have all been reddened to our adopted $A_V$ = 1.5 mag and using the \citet{Whittet_ExtinctionCurve_2004ApJ} extinction curve, which has a weaker 2200 \AA\ bump than the commonly used \citet{cardelli_relationship_1989} curve. The effect of the bump can be seen as an undulation from 2000 \AA\ to 2800 \AA\ in the otherwise featureless reddened excess component blackbody spectrum.}
    \label{fig:model}
\end{figure*}

The full STIS UV spectrum of FU Ori is shown in Figure \ref{fig:model}. We include the binned continuum points derived from the COS spectrum, which are in good agreement with the continuum level of the STIS G140L spectrum. The continuum points from the 1800 \AA\ to 2400 \AA\ region of the STIS G230L spectrum used for our fit to the FUV excess are also marked. 


The three bright lines in the G230L spectrum are the \ion{C}{2}] 2335 \AA, \ion{Fe}{2} 2507/2509 \AA, and \ion{Mg}{2} 2795/2802 \AA\ features, respectively. We further discuss their appearance in Appendix \ref{sec:NUVMeas}.

\section{Fitting the FUV Excess Emission} \label{sec:cont_fit}
We identify the NUV/FUV excess emission by comparing the observed spectrum to a viscous gas accretion disk model spectrum. Our model is described in detail in \citet{Carvalho_V960MonPhotometry_2023ApJ} but we summarize it here. 

We assume that for $r < 150 \ R_\odot$ (43 $R_*$), the FU Ori accretion disk is thin \citep[$H/R < 0.01$,][]{Zhu_outburst_FUOri_2020MNRAS} and viscously heated. We then adopt a modified version of the \citet{Shakura_sunyaev_alpha_1973A&A} $\alpha$-disk temperature profile, 
\begin{equation} \label{eq:TProf}
    T^4_\mathrm{eff}(r) = \frac{3 G M_* \dot{M}}{8 \pi \sigma r^3} \left( 1 - \sqrt{\frac{R_\mathrm{inner}}{r}}  \right)    ,
\end{equation}
where $R_\mathrm{inner}$ is the inner radius of the accretion disk, $M_*$ is the mass of the central star, $\dot{M}$ is the stellar accretion rate, $G$ is the gravitational constant, and $\sigma$ is the Stefan-Boltzmann constant. We assume that for $r < \frac{49}{36} R_\mathrm{inner}$, $T_\mathrm{eff}(\frac{49}{36}R_\mathrm{inner}) = T_\mathrm{max}$ \citep{Kenyon_FUOri_disks_1988ApJ}. 

We populate the annuli of the disk using PHOENIX model stellar spectra \citep{Husser_Phoenix_2013A&A} with log$g=1.5$ and the appropriate $T_\mathrm{eff}(r)$. For the FU Ori system parameters, we adopt most of those reported in \citet{Zhu_outburst_FUOri_2020MNRAS}: $M_* = 0.6 \ M_\odot$, $R_\mathrm{inner} = 3.52 \ R_\odot$, $R_\mathrm{outer} = 0.7$ AU. For the distance to the system, we use the Gaia-derived distance of 404 pc to the $\lambda$ Ori cluster, of which FU Ori is a member \citep{Kounkel_LamOriDist_2018AJ,
Roychowdhury_FUOriV883OriDist_2024RNAAS}. We assume an inclination of 35$^\circ$ \citep{Perez_FUOriALMA_2020ApJ}. We initially explored models across a range of $\dot{M}$ and $A_V$ (described in Appendix \ref{app:MdotAvTest}) and found that $\dot{M} = 10^{-4.49} \ M_\odot$ yr $^{-1}$ is a better fit to the NUV spectrum than the $\dot{M} = 10^{-4.42} \ M_\odot \ \mathrm{yr}^{-1}$ used by \citet{Zhu_outburst_FUOri_2020MNRAS}. This analysis also showed that the $A_V$ is weakly constrained but agrees with the \citet{Zhu_outburst_FUOri_2020MNRAS} value of $A_V = 1.5$ mag. Therefore, in our fiducial disk model we use $\dot{M} = 10^{-4.49} \ M_\odot$ yr $^{-1}$ and $A_V = 1.5$ mag.

For our extinction correction, we adopt the \citet{Whittet_ExtinctionCurve_2004ApJ} extinction curve, which was based on stars located behind the Taurus Molecular Cloud. This curve has a much weaker 2175 \AA\ "bump" than the typically-used \citet{cardelli_relationship_1989} and has been found to be a better fit to the ISM conditions in star forming regions. In our STIS spectrum, the \citet{cardelli_relationship_1989} curve over-predicts the 2175 \AA\ extinction. For simplicity, we fix $R_V = 3.1$. 

The STIS spectrum is well matched by the viscous disk component of our model until the sharp continuum break at $\sim 2060$ \AA. The break is due to the \ion{Al}{1} continuum opacity, which has a jump in that region that grows stronger with temperature for $T_\mathrm{eff} > 5000$ K \citep{TravisMatsushima_Opacities_1968ApJ}. The \ion{Al}{1} continuum opacity may be overestimated in the model atmospheres, or it is possible that in this innermost region of the disk our assumption of an LTE plane-parallel atmosphere no longer holds. Regardless, the agreement between the continuum level of the disk model and the continuum in the $2100-3100$ \AA\ range covered by the E230M spectrum confirms that the disk model does not require $T_\mathrm{max} \sim 9000$ K to match the observed flux \citep{hartmann_FUOriProfile_2011arXiv1106}. 

\begin{figure}
    \centering
    \includegraphics[width=\linewidth]{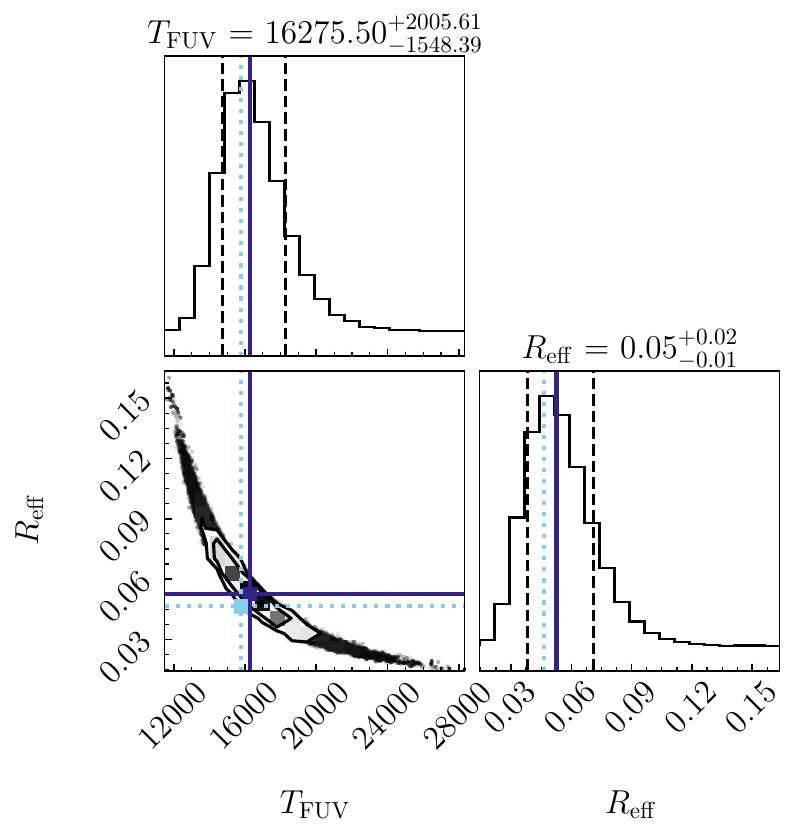}
    \caption{The $\mathtt{corner}$ plot for our MCMC fit for the FU Ori FUV excess. Our adopted best-fit values are marked by the navy vertical lines, showing the medians of the histograms. The modal values, marked by light blue dotted lines, are identical to the median values. The contours in the 2-d histogram mark the 0.5, 1.0, 1.5, and 2.0 $\sigma$ levels, while the vertical dashed lines mark the 16th and 84th percentiles in the 1-d histograms. }
    \label{fig:corner}
\end{figure}

Blueward of 2100 \AA, the observed FUV continuum exceeds the disk model spectrum by several orders of magnitude, requiring the an additional component in the model to match the data. We model the component as a simple Planck function of a single temperature, $T_\mathrm{FUV}$ having an effective radius $R_\mathrm{eff}$. The additional flux is $F_\mathrm{BL} = \pi B_\lambda(T_\mathrm{FUV}) \left( \frac{R_\mathrm{eff}}{d} \right)^2$, where $d = 404$ pc. To determine the best-fit $R_\mathrm{eff}$ and $T_\mathrm{FUV}$ of the emitting region, we use Markov-Chain Monte-Carlo ensemble sampling from the $\mathtt{emcee}$ package \citep{FM_emcee_2013PASP}. We summarize the posterior distributions for the $T_\mathrm{FUV}$ and $R_\mathrm{eff}$ of the emission region in Figure \ref{fig:corner}, constructed using the $\mathtt{corner}$ \citep{corner_FM_2016} package. 

Although we considered multi-temperature models, we ultimately find that a single-temperature component provides the best fit to the spectrum. Introducing even one additional emission component with a different temperature results in a bimodal posterior distribution with one peak at $T_1 \sim 6000$ K, essentially the $T_\mathrm{max}$ in our disk model, and another at $T_2 \sim 14000 - 16000$ K, the same as the $T_\mathrm{FUV}$ of the one-temperature solution. 

We also considered the commonly-used hydrogen slab model \citep{valenti_tts_1993}. However, as mentioned above, $T > 12,000$ K is necessary to explain the shape of the FUV continuum we detect. At these temperatures, hydrogen is almost entirely ionized, and the continuous opacity is dominated by emission from metals, particularly C, Si, Mg, and Al \citep{TravisMatsushima_Opacities_1968ApJ}. Since we do not see the distinct jumps at the wavelengths where each particular species opacity should dominate (e.g., 2070 \AA\ for Al), we can assume that the emission is largely blackbody.

The best-fit parameters for our single temperature model are $T_\mathrm{FUV} = 16275^{+2005}_{-1548}$ K and $R_\mathrm{eff} = 0.05^{+0.02}_{-0.01} \ R_\odot$. These values are robust to allowing $A_V$ and $\dot{M}$ to vary in the model, as we demonstrate in Appendix \ref{app:MdotAvTest}. Our best fit model\footnote{In the 1700-1800 \AA\ region, the model appears to over-estimate the continuum. However, this is near the detector cut-off at the reddest points of the COS spectrum and it is likely that the uncertainties on those photometric points are underestimated. The detected continuum in that wavelength range should thus be considered a lower bound.}
is shown along with the spectra in Figure \ref{fig:model}.

If we assume that the geometry of the emission region is a disk or torus at the equator the $R_\mathrm{eff}$ corresponds to an annulus with $R_\mathrm{outer} = 3.52 \ R_\odot$ and a $\Delta r = 4\times 10^{-4} \ R_\odot$. We interpret this extremely narrow width as evidence of an accretion shock at the disk-star boundary due to the disk surface accretion flow impacting the stellar surface.

Continuing to assume a disk-like geometry for the emission region, the FUV excess luminosity, $L_\mathrm{FUV}$, is given by $L_\mathrm{FUV} = \frac{2\pi}{\cos i} R_\mathrm{eff}^2 \sigma T_\mathrm{FUV}^4 = 0.11 \ L_\odot$. Considering only the projected area we observe, instead of a disk-like geometry, $L_\mathrm{FUV} = \pi R_\mathrm{eff}^2 \sigma T_\mathrm{FUV}^4 = 0.04 \ L_\odot$. We compare those values with the CTTS population in Section \ref{sec:CTTSComp}.

The fiducial model that we adopt is distinct from previously proposed boundary layer models that modify the temperature profile of the disk assuming sheer between the disk and star contributes to excess heating \citep[e.g.,][]{Regev_diskStarBoundaryLayer_C_1983A&A,Popham_boundaryLayersInPMSDisks_1993ApJ, Kley_FUOriBL_J_1996ApJ}. We explored such modified temperature profiles and found that they were insufficient to reproduce the full UV spectrum. Our results of these tests are presented in Section \ref{sec:PreviousModels}.

\section{Discussion} \label{sec:discussion}
The observed FUV emission for FU Ori exceeds the viscous disk model by a factor of $>10^4$ (see Figure \ref{fig:model}), indicating the presence of a bright emission source at the disk-star interface. In this Section, we compare our results with previous theoretical predictions of a boundary layer in the FU Ori system. We then compare our measured $T_\mathrm{FUV}$ with estimates of boundary layer properties based on analyses of the 2001 STIS E230M spectrum. We also discuss other potential mechanisms for the FUV emission and why we disfavor them in our interpretation.

\subsection{A shock at the star-disk interface}\label{sec:shockDescription}

Our best-fit model for the FUV continuum excess is of an emission source with a single temperature of 16,000 K and a filling factor of 0.02\% on the surface of the star. We also constrain the $\alpha$-disk component of the system to have a $T_\mathrm{max} =$ 5,800 K, a bit cooler than, but consistent with, the $T_\mathrm{max} = $6,050 K reported by \citet{Zhu_outburst_FUOri_2020MNRAS}. Notably, we do not find evidence in the spectrum from components with temperatures in between 5,800 K and 16,000 K as would be expected from shear-heated boundary layer models. 

Given the small filling factor of the 16,000 K component relative to the $r \sim R_\mathrm{inner}$ region of the disk or the stellar surface, which is presumably also at $R_* \sim R_\mathrm{inner}$, and the sudden jump in temperature, we conclude that the emission arises from a shock at the star-disk interface. 

In previous magneto-hydro-dynamical simulations of accretion disks vertically threaded by magnetic fields \citep[like we expect to be the case in FU Ori,][]{Donati_FUOri_2005Natur}, the accretion flow does not occur along the midplane but the radially inward mass flux is concentrated in a vertically thin region above the disk photosphere \citep{gammie_surfaceAccretion_1996ApJ, Zhu_Stone_SurfaceAccretion_2018ApJ}. This accretion geometry is termed ``surface accretion" \citep{Zhu_outburst_FUOri_2020MNRAS} and the shock we observe is at the boundary where the supersonic surface accretion flow, with velocity $v_\mathrm{inflow}$, collides with the stellar photosphere.

Assuming the temperature of the material is near the $T_\mathrm{max}$ of the system implies a Mach number of $v_\mathrm{inflow}/c_s \sim 4$. Therefore, when the supersonic flow reaches the stellar surface, it should produce a strong shock. The temperature of the shocked material can be then computed using the typical Hugoniot strong shock conditions,
\begin{equation}
    T_\mathrm{shock} = \frac{3}{16}\frac{\mu m_H}{ k_B} v_\mathrm{inflow}^2,
\end{equation}
where $\mu = 0.5$ is the mean molecular weight assuming the gas is dominated by ionized hydrogen, and $m_H$ is the proton mass. If we use an inflow velocity of $38$ km s$^{-1}$, then $T_\mathrm{shock} \approx 16,000$ K. This is consistent with the $v_\mathrm{inflow} = 20-40$ km s$^{-1}$ of the surface accretion layer seen in simulations of the FU Ori accretion disk \citep{Zhu_outburst_FUOri_2020MNRAS}.

\subsection{Is this emission consistent with the presence of a boundary layer?}\label{sec:BLDiscussion}
While we interpret the FUV emission as arising from a shock, here we consider the necessary properties of a boundary layer that might produce the FUV emission. We will first discuss the nature of boundary layers in the context of YSO accretion disks and compare predictions of such boundary layers with our observed spectra. In the following sections, we turn toward specific boundary layer models and directly contrast those with the data. 

In the work by \citet{Lynden-Bell_Pringle_ViscousDisks_1974MNRAS}, \citet{Pringle_accretionDisksReview_1981ARA&A}, and \citet{Popham_boundaryLayersInPMSDisks_1993ApJ}, the boundary layer is a region where material that is moving at Keplerian orbit velocities of $200 - 300$ km s$^{-1}$ at $r \approx R_\mathrm{inner}$ slows to the stellar rotational velocity of $10 - 30$ km s$^{-1}$ \citep{Nofi_rotational_2021ApJ}. The energy released by the shear is expected to heat the gas nearest the star sufficiently that $L_\mathrm{BL} \sim L_\mathrm{acc}$. The region is also predicted to have a width comparable to the scale height of the disk at $R_\mathrm{inner}$, or $dR_\mathrm{BL} = 0.1 \ R_\mathrm{inner}$. 

Compared with the observed FUV emission, which has a luminosity of $0.1 \ L_\odot$, the continuum is more than 700 times underluminous for a shear-heated boundary layer. The expected total area of a boundary layer of thickness $0.1 \ R_\mathrm{inner}$ is 880 times greater than that which we find for our best-fit model. Matching both the size and luminosity of the boundary layer predicted for an FU Ori system would require that the FUV emission source be obscured by an $A_V > 6$ mag. This may be the case if a majority of the boundary layer is buried by cooler upper layers of the disk atmosphere that trap the hot photons in the innermost region of the disk.  

It is possible to match only the size of a boundary layer by treating the source of the FUV continuum as optically thin, rather than the optically thick model which we estimate to have $dR = 4 \times 10^{-4} \ R_\odot$. Then, the expected $dR_\mathrm{BL}$ of would imply an optical depth of $\tau = -\ln \left(1 - 0.0004/0.352 \right) < 0.001$. Given the high density expected in this region \citep[$n > 10^{15}$ cm$^{-3}$,][]{Zhu_outburst_FUOri_2020MNRAS}, the emission is unlikely to be optically thin.

Another prediction in the boundary layer model is that there should be a continuous temperature increase as $r \rightarrow R_\mathrm{inner}$, rather than the sharp jump we find with our single temperature model. To investigate the potential for a continuous temperature model to fit the FUV excess, we tested using a temperature profile with a power law dependence on the area of the emission region. Then the flux of such a boundary layer would be given by $F_\mathrm{FUV} = \pi \sum_i^{N_T} B_\lambda(T_i)(A_i/d^2)$, where $N_T=10$ is the total number of temperature components we used and $A_i$ is the area of each component of temperature $T_i$. The power law is given by $A_i = A_0 (T_0/T_i)^\gamma$, where $A_0$ is the reference area for the lowest temperature component, $T_0$. 

We again performed an MCMC fit and found that $T_0 = 6400 \pm 1000$ K, $A_0 = 0.04 \pm 0.03$, and $\gamma = 5.03 \pm 0.8$. The total area of the boundary layer integrated over the 10 temperature components is 0.4 $R_\odot^2$, which would correspond to an annulus of $dR_\mathrm{BL} = 0.017 \ R_\odot = 0.005 \ R_\mathrm{inner}$. Even allowing a continuous temperature increase, the preferred model yields a thin boundary layer (or one that requires $\tau << 1$) and a steep dependence on the size of each emitting region. The boundary layer here is not much thicker than is expected for our single-temperature model, implying it is either much thinner than expected for FU Ori systems or that the FUV continuum is not produced by such a boundary layer.

\subsection{Comparison of the shock model with previous boundary layer models}\label{sec:PreviousModels}

We find good agreement between the thin viscous accretion disk $+$ blackbody shock model and the observed FUV spectrum. Moving towards the NUV, the component of the spectrum that remains relatively dominated by the viscous accretion disk continuum is consistent with the modified \citet{Shakura_sunyaev_alpha_1973A&A} model we adopt, including the isothermal $T(r < \frac{49}{36} R_\mathrm{inner}) = T_\mathrm{max}$ assumption. 

With our FUV spectrum, we can also test boundary layer models wherein the sheer between the star and disk produce a boundary layer with a temperature dependent on the efficiency of angular momentum accretion, $j$, of the star \citep{Popham_boundaryLayersInPMSDisks_1993ApJ, Popham_boundaryLayerSpectraLineProfiles_1996ApJ}. As a result of the increased heating from less efficient angular momentum transport, the \citet{Shakura_sunyaev_alpha_1973A&A} temperature profile in the thin disk case becomes,
\begin{equation} \label{eq:TProfAlt}
    T^4_\mathrm{eff}(r) = \frac{3 G M_* \dot{M}}{8 \pi \sigma r^3} \left( 1 - j\sqrt{\frac{R_\mathrm{inner}}{r}}  \right)    ,
\end{equation}
where $j \equiv \dot{J}/\dot{M}\Omega(R_*)R_*^2$, $\dot{J}$ is the angular momentum transfer rate onto the star, $\Omega(R_*)$ is the Keplerian frequency at the stellar surface and the new location of $T_\mathrm{max}$ in the disk is at $r = \frac{49}{36}j^2 R_\mathrm{inner}$. 

We tested varying the $j$ parameter in our model and found that it was inadequate to simultaneously reproduce the visible, NUV, and FUV spectra. If we allow $j$ to vary and include a shock component with a variable $T_\mathrm{FUV}$ and $R_\mathrm{eff}$, the fit converges to $j \sim 0.95$ and T$_\mathrm{eff}$ and $R_\mathrm{eff}$ values close to those in our fiducial model. The $j \sim 0.95$ temperature profile is almost isothermal for $r < \frac{49}{36}j^2 R_\mathrm{inner}$ and produces an SED that is indistinguishable from our fiducial model temperature profile. 

We note that our model relies on the thin disk approximation, which may break down in the shear-heated region closest to the central star. The models in \citet{Popham_boundaryLayersInPMSDisks_1993ApJ} account for changes to the disk scale height due to the excess heating, which can produce different temperature profiles at $r \sim R_\mathrm{inner}$ than the thin disk approximation. For instance, in all of their temperature profiles, the $T(r < 1.2 \ R_\mathrm{inner}) < T_\mathrm{max}$, regardless of choice of $j$. In the thin disk case, this is only true for $j > 0.9$. Furthermore, the line profiles predicted for $j < 0.6$ reproduce the flat-bottomed square line profiles observed in the visible and NIR spectra of FU Ori \citep{Kenyon_FUOri_disks_1988ApJ, PetrovHerbig_FUOriLineStructure_2008AJ, Zhu_FUOriDifferentialRotation_2009ApJ} best, though the thin disk models with $j$ values in this range are inconsistent with the SED from the NUV to the NIR. 



Another potential boundary layer model for FU Ori objects may be found in the accreting compact object literature. Existing white dwarf accretion disk boundary layer models could be good analogs for the very narrow, hot boundary layer that we attribute to a shock \citep{Hertfelder_CompactObjectBLs_VerticalStructure_2017A&A}. In these models, in the final $r < 1.01 \ R_\mathrm{inner}$ of the disk, the temperature of the boundary layer increases rapidly to $5-10 \times$ the temperature at $r \sim 1.1 \ R_\mathrm{inner}$. The boundary layer typically has a thickness of $\Delta r < 0.006 \ R_\mathrm{inner}$ and spread upward along the surface of the compact object, reaching latitudes of 35$^\circ$. This is a factor of 10 greater than our limit on the thickness of a toroidal boundary layer region around FU Ori, but if such a boundary layer were present in this system, we can compare the measured $T_{BL}$ with a prediction based on the white dwarf case.

As a crude estimate of how the boundary layer temperature from \citet{Hertfelder_CompactObjectBLs_VerticalStructure_2017A&A} might scale going from a white dwarf system to an FU Ori system, we can use the fact that $T_\mathrm{max}\propto (M_*\dot{M}/R_\mathrm{inner}^3)^{1/4}$. Scaling then from the white dwarf parameters adopted by \citet{Hertfelder_CompactObjectBLs_VerticalStructure_2017A&A}, we get
\begin{gather*}
    T_\mathrm{BL, FU} = 350\times 10^3 \ \mathrm{K} \ \left( \frac{0.6 \ M_\odot}{1.0 \ M_\odot} \right)^{0.25} \times \\ \left( \frac{10^{-4.49} \ M_\odot \mathrm{yr}^{-1}}{10^{-8} \ M_\odot \mathrm{yr}^{-1}} \right)^{0.25} \left( \frac{3.52 \ R_\odot}{5.6 \times 10^8 \mathrm{cm}} \right)^{-0.75} = 24,000 \ \mathrm{K},
\end{gather*}
which is 50\% larger than the $T_\mathrm{FUV} \approx 16,000$ K we find. This is a simple scaling exercise, but demonstrates that boundary layer models based on compact object accretion disks may be applicable to the FU Ori disk and should be explored further.

If we consider a geometry similar to that of the white dwarf boundary layer described above, where the emission region is a rectangular band on the surface of the star with a latitudinal size of $h$ and wraps around the equator of the star, then its area projected along our line of sight is given by $2 \pi R_* h \sin i$. Requiring that the observed area be equal to $\pi R_\mathrm{eff}^2$ and solving for $h$ in this case yields a similar value to the annular emission geometry: $h = 6 \times 10^{-4} \ R_\odot$. In this geometry, it is likely that we are only seeing a very small part of the total emission surface, and that the rest is obscured by the dense disk midplane.

Ultimately, we believe our proposed accretion shock interpretation to be the most reasonable explanation for our observations. The interpretation is complementary to what may be expected from the flows predicted by the radiation MHD models in \citet{Zhu_outburst_FUOri_2020MNRAS}. 

\subsection{Excluding magnetic activity as a source of FUV continuum emission} \label{sec:magAct}

A potential source of FUV continuum emission is magnetic activity, particularly in events like stellar flares. In FU Ori objects, the X-ray emission has been established as likely arising from heightened magnetic activity due to the extreme temperatures of the X-rays \citep{kuhn_comparison_2019}. Here, we consider whether the FUV emission might arise from a transition region between the star and the X-ray emitting corona or from serendipitous flare-like activity during our observation.

Studies of superflares in magnetically active M dwarfs, known to have $\sim$ kG field strengths like YSOs, describe the FUV continuum during flares as 15,000 - 30,000 K blackbodies, with significant enhancement in several high temperature emission lines \citep{Kowalksi_StellarFlares_Review_2024LRSP}. While we find a continuum temperature at the lower end of this range, and detect common features like the \ion{C}{4} and \ion{Si}{4} doublets and \ion{He}{2} 1640 \AA, there are other bright emission lines seen in flare spectra that we at best marginally detect \citep[e.g., \ion{C}{3} 1175, \ion{Si}{3} 1206.51, the \ion{N}{5} doublet, and the \ion{C}{1} multiplet,][]{Lloyd_FUVFlaresOnMDwarfs_2018ApJ, MacGregor_ProxCenFlaresInMM_FUV_2021ApJ}. 

Although the continuum may be similar to that observed in flares, the brightness and persistence of the emission would imply extraordinary flare activity. Integrating the FUV luminosity of $0.1 \ L_\odot$ over the 8482 s of exposure time gives a total emitted energy of $3.5 \times 10^{36}$ erg. This is comparable to the \textit{bolometric} energy of the most powerful superflares observed in flare stars \citep{Simoes_FlareStarsTemperaturesContinuum_2024MNRAS}. The FUV flux level is also constant during the four COS exposures. For flares to have produced the emission would require 4 almost identical, extremely powerful flares to have occurred consecutively during the time of observation, each lasting no longer than the time of an individual exposure. 

It may be that the high temperature lines like the \ion{C}{4} and \ion{Si}{4} doublets and \ion{He}{2} 1640 \AA\ feature are emitted from a transition region between the star-disk interface and the corona from which the X-rays are emitted. However, it is unlikely that the FUV continuum is produced by magnetic activity. This is supported by the stability of the FUV continuum emission and the fact that while coronal line emission is observed in magnetically active stars, bright continuum emission is only seen during flares. 

\subsection{Excluding 2-photon emission as the source of FUV emission}\label{sec:OtherModels}
In addition to multi-component models and variations on the viscous disk temperature profile, we considered the hydrogen two-photon process as a source of the FUV emission. We expect that the number densities associated with the process ($n_e < 10^3$ cm$^{-3}$) are much lower than those predicted for the region near the star in the FU Ori system \citep{Zhu_outburst_FUOri_2020MNRAS}. Therefore, if the FUV emission were indeed two-photon, it would instead suggest a nebular origin, rather than a boundary layer or shock. 

To test the possibility that we are seeing two-photon emission, we substitute the  Planck function in the model with the two-photon model given in \citet{France_2photon_2011ApJ}. Following the procedure in Section \ref{sec:cont_fit}, the best-fit $R_\mathrm{eff} = 0.2 \ R_\odot$ and the hydrogen column density is $3 \times 10^{21}$ cm$^{-2}$. 

Although the model fits the FUV continuum well for $1220$ \AA\ $< \lambda < 1600$ \AA, the flux at $1800$ \AA\ $< \lambda < 2200$ \AA\ is insufficient to match the observed spectrum. Analyzing the count rate in the COS spectrum, we also detect the continuum blueward of $\lambda = 1200$ \AA\ at a $10\sigma$ level (see Appendix \ref{app:COSCont}). The spectrum of the two-photon process has a sharp cutoff at $\lambda < 1215$ \AA\ and would therefore not be detected. This rules out the two-photon process as the source of the FUV continuum.

\subsection{The UV luminosity of FU Ori in context} \label{sec:CTTSComp}

With our two assumptions about the geometry of the FUV emission source (isotropic versus disk-like), we estimate the FUV luminosity of FU Ori to be $L_\mathrm{FUV} = 0.04-0.11 \ L_\odot$. The ratio of FUV luminosity to accretion luminosity is $L_\mathrm{FUV}/L_\mathrm{disk} = 0.5-1.2 \times 10^{-3}$. Considering the $L_\mathrm{X} = 1.6 \times 10^{-3} \ L_\odot$, the $L_\mathrm{X}/L_\mathrm{disk} = 2.5 \times 10^{-5}$ \citep{kuhn_comparison_2019}. 

To better contextualize these values relative to magnetospherically accreting young stars, we compare the continuum (i.e., excluding Ly$\alpha$) $L_\mathrm{FUV}$ and $L_\mathrm{X}$ for FU Ori with CTTSs in Taurus using data from \citep{yang_HST_TTS_FUV_Survey_2012ApJ}. For the median CTTS in the sample, $L_\mathrm{FUV} = 10^{-2.7} \ L_\odot$. FU Ori is more FUV-luminous than the median object in the CTTS sample by almost a factor of 100. FU Ori is also much more X-ray luminous than typical members of the CTTS sample. The objects in this parameter space that have comparable FUV and X-ray luminosities to FU Ori are K0 or G type stars and up to 2$\times$ more massive than FU Ori. 

The increased disk luminosity during an FU Ori outburst is predicted to drive disk chemistry by sublimating volatiles on icy dust grains, leading to enrichment of complex organic molecules \citep{Molyarova_ChemicalSignaturesFUOriOutbursts_2018ApJ}. The increased hard radiation from FU Ori in outburst may also have significant implications for disk chemistry during outbursts. If we calculate the FUV energy density, $u_{FUV}(10 \ \mathrm{au}) = \frac{1}{4\pi c}\sigma_{SB}T_\mathrm{eff}^4 \left(\frac{0.05 \ R_\odot}{10 \ \mathrm{AU}} \right)^2 = 1.15 \times 10^{-8}$ erg cm$^{-3}$, or $10^{5.3} \ G_0$. Closer-in, $u_\mathrm{FUV}(1 \ \mathrm{AU}) \sim 10^7 \ G_0$. As expected from the larger $L_\mathrm{FUV}$, this is a factor of 10-100 greater than the typical continuum $u_\mathrm{FUV}$ near the central star in a CTTS system \citep{France_UVRadiationFields_2014ApJ}. It is also greater than the FUV energy density from the Orion Trapezium Cluster found to drive a photoevaporative disk outflow in nearby YSO \citep{Berne_OutflowDrivenbyFUV_2024Sci}. The total $u_\mathrm{FUV}$ can be even greater if Ly$\alpha$ is accounted for in the system, which in CTTSs can contribute $10\times$ the continuum FUV flux \citep{France_UVRadiationFields_2014ApJ}. Studies considering the effects of FU Ori outbursts on the chemical environment in protoplanetary disks should take into account both the increased $L_\mathrm{bol}$ and $L_\mathrm{FUV}$.

\begin{figure}
    \centering
    \includegraphics[width = \linewidth]{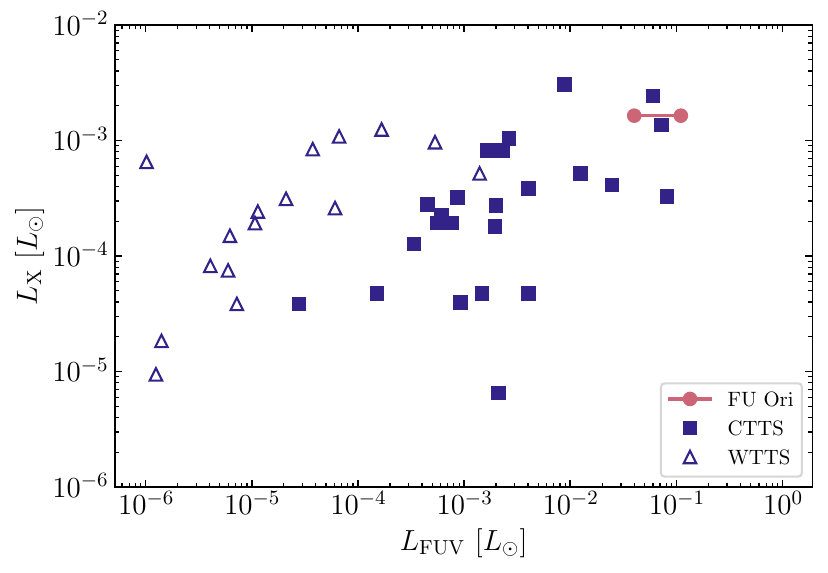}
    \caption{The $L_\mathrm{FUV}$ (this work) and $L_X$ \citep{kuhn_comparison_2019} for FU Ori, compared with YSOs in Taurus \citep{yang_HST_TTS_FUV_Survey_2012ApJ}. FU Ori has among the highest $L_X$ and has the highest $L_\mathrm{FUV}$ of the sample. The range of $L_\mathrm{FUV}$ for FU Ori reflects the two values we calculate depending on the assumed geometry of the emission region. The sources that have $L_\mathrm{FUV} > 0.05 \ L_\odot$ are T Tau, HD 143006, and RY Tau, which are much earlier spectral types than the FU Ori central object would be \citep[K1, G3, and G0, respectively,][]{herczeg_survey_2014}.}
    \label{fig:Lx_vs_LUV}
\end{figure}

\section{Conclusions} \label{sec:conclusions}
The HST COS/STIS spectra of FU Ori contain FUV emission exceeding by a factor of $> 10^4$ at 1400 \AA the flux predicted by the viscous disk accretion model that explains in full the $4000-22,000$ \AA\ wavelength range. To understand the nature of the excess, we isolated the continuum emission in the COS spectrum. We carefully masked out several regions that either showed clear line emission or are known to show line emission in accreting YSOs. We then fit the FUV spectrum with a viscous disk model combined with a blackbody spectrum, varying the $T_\mathrm{FUV}$ of the spectrum and the $R_\mathrm{eff}$ of the emitting region. Our combined model shows that the temperature and effective radius of the emitting region are $T_\mathrm{FUV} = 16,000 \pm 2,000 \ \mathrm{K}$ and $R_\mathrm{eff} = 0.05 \pm 0.02 \ R_\odot$. 

The measured temperature and the size of the FUV emission region are consistent with expectations for a shock at the disk-star boundary. The shock arises from the collision of the highly supersonic disk surface accretion flow with the stellar photosphere. The FUV continuum luminosity is $L_\mathrm{FUV} = 0.04-0.11 \ L_\odot$, which is $\sim 100 \times$ greater than typical values for CTTSs. The resulting FUV energy density is $10^7 \ G_0$ at 1 AU and $10^{5.3} \ G_0$ at 10 AU, comparable with the most extreme FUV interstellar radiation fields. The large increase in FUV flux from the inner disk could drive photochemistry in a larger portion of the outer disk relative to quiescent systems. 

We encourage detailed theoretical modeling of the innermost $r < 0.1$ AU of the FU Ori system at outburst, incorporating a disk-star boundary layer in order to better understand the gas dynamics in the region. In future work, we will present an analysis of the many emission lines in the COS spectrum to understand their origin and the properties of the massive, high velocity outflows near the star.

\section{Acknowledgements}
This research was supported, in part, by grant HST-GO-17176.001-A from STScI. The authors thank Lee Hartmann for his insightful comments on the analysis and interpretation of the data. 
The authors acknowledge Will Fischer, for many things over the years, but most relevant here is the volunteering of valuable, unsolicited advice regarding our $cenwave$ choice.

\bibliography{references}{}
\bibliographystyle{aasjournal}

\appendix 

\restartappendixnumbering

\section{Line Profiles of Common CTTS FUV Emission Features}\label{sec:lineProfs}

\begin{figure*}[!htb]
    \centering
    \includegraphics[width = 0.98\linewidth]{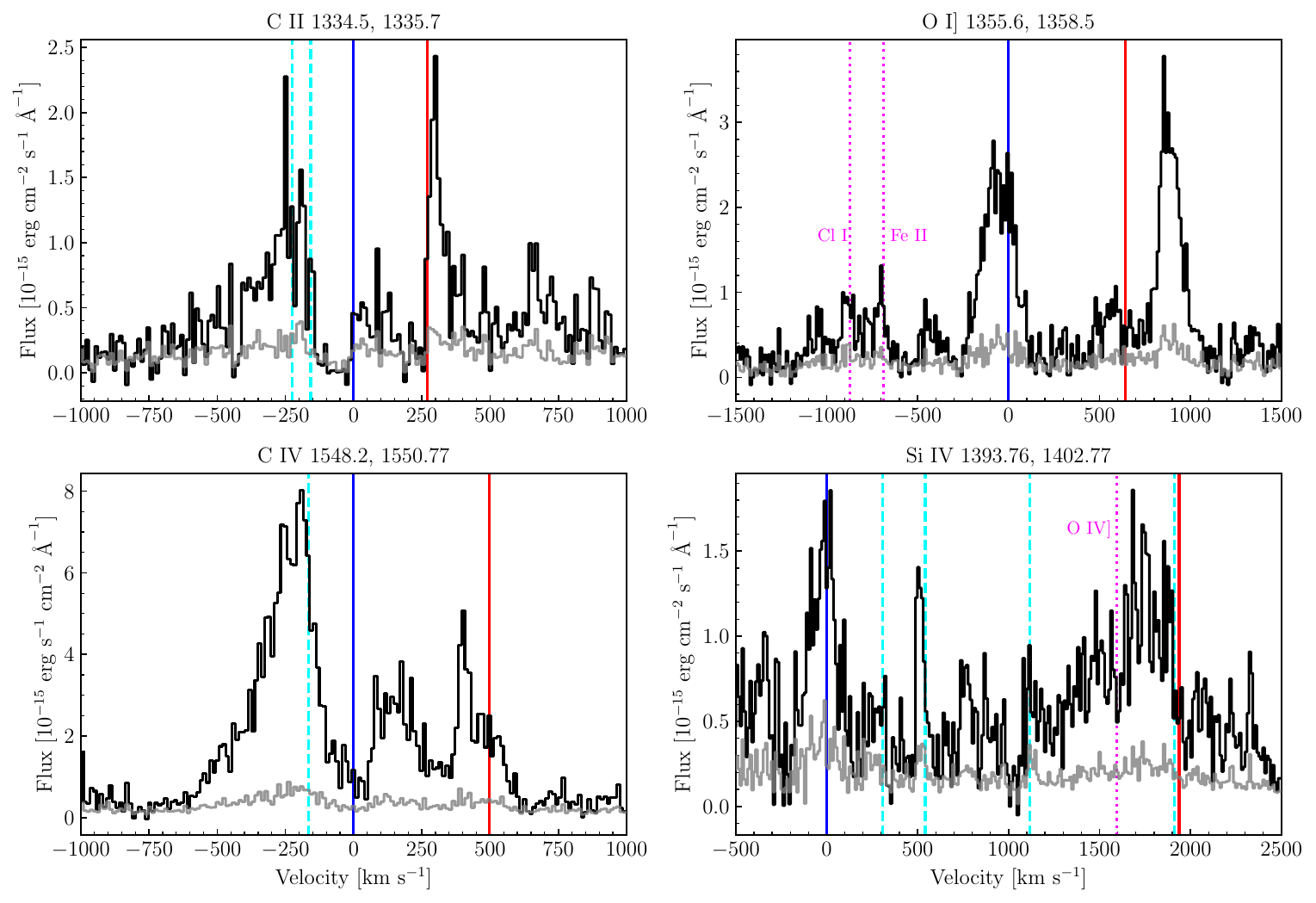}
    \caption{Selected bright emission line doublets commonly studied in the FUV spectra of YSOs. The data are plotted in black and the uncertainty on each point is shown in light grey. The blue and red lines mark the blue and red 
    components of the doublets. The vertical cyan dashed lines mark the rest-wavelengths of H$_2$ lines and the vertical magenta dotted lines mark the rest frequencies of known features that may be contaminating the doublet line profiles (Cl I 1351.657, Fe II 1352.487, and O IV] 1401.17).}
    \label{fig:COSDoubletLines}
\end{figure*}

\begin{figure*}[!htb]
    \centering
    \includegraphics[width = 0.98\linewidth]{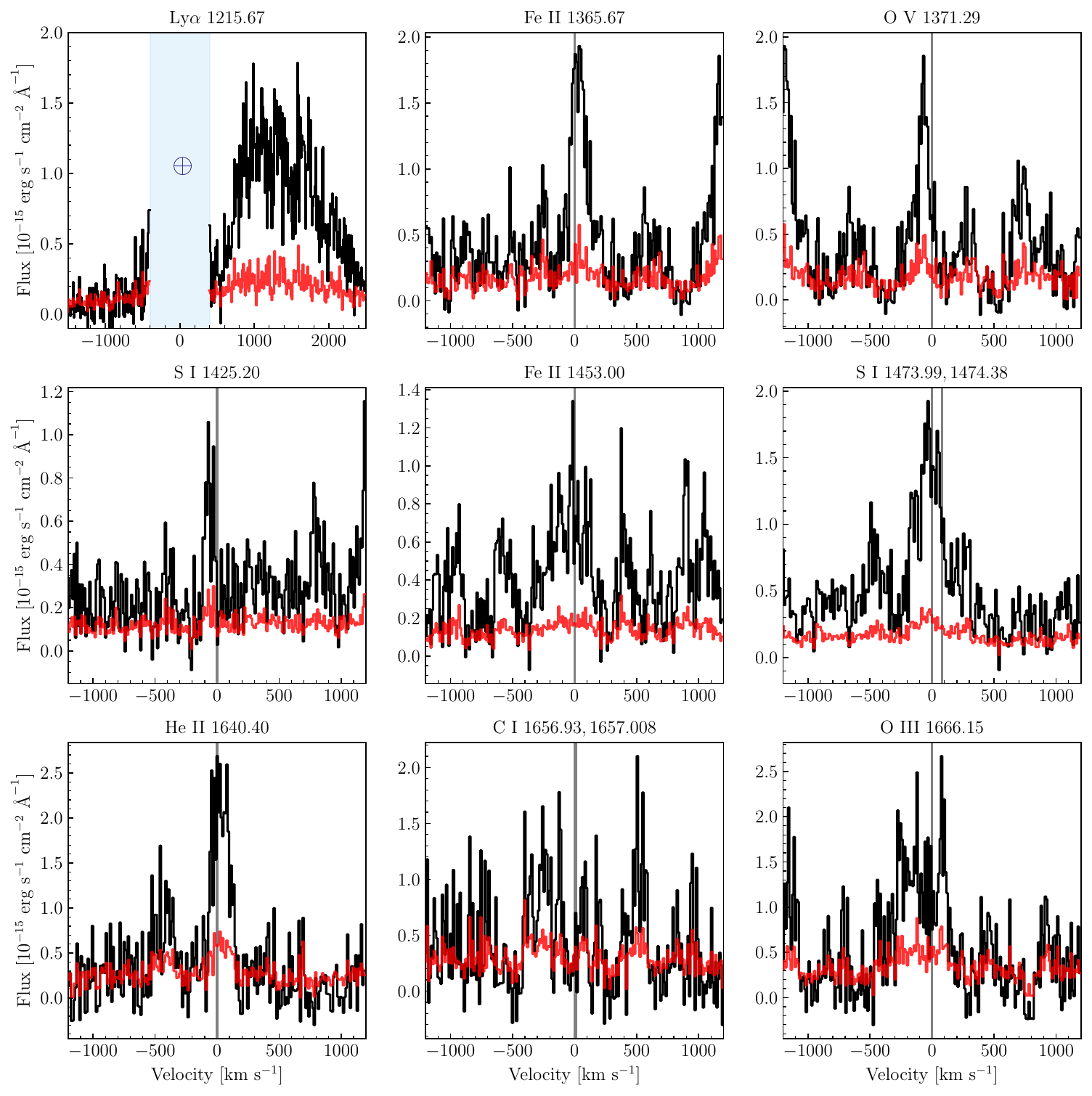}
    \caption{Detected bright emission lines commonly seen in the FUV spectra of YSOs. The data are plotted in black and the uncertainty on each point is shown in red. The rest-frequency of each line marked by the vertical grey line. Notice that, if present, the \ion{C}{1} 1656 feature is only marginally detected.}
    \label{fig:COSLines}
\end{figure*}

The COS spectrum shown in Section \ref{sec:data} has several bright emission lines, many of which have profiles similar to those seen in CTTS systems with accretion and strong outflows, wherein the lines are highly self-absorbed. The brightest features in the COS spectrum are Ly$\alpha$, the \ion{C}{2} 1334.5/1335.7 doublet, the \ion{C}{4} 1548.2/1550.77 doublet, the \ion{Si}{4} 1393/1402 doublet, and \ion{He}{2} 1640.4 (Figure \ref{fig:COSDoubletLines}, \ref{fig:COSLines}). In this Appendix, we discuss the lines shown in Figures \ref{fig:COSDoubletLines} and \ref{fig:COSLines} and compare them with their counterparts in quiescent young stars. A detailed analysis of the lines will be presented in a future paper. 

There is almost no \ion{C}{1} detected in the spectrum. We marginally detect the features at 1656/1657, which at $T \sim 16,000$ K (according to the CHIANTI atomic line database, assuming Local Thermodynamic Equilibrium; LTE) are predicted to be the brightest \ion{C}{1} features \citep{dere_chianti_1997A&AS, delZanna_ChiantiV10_2021ApJ}. It is possible that the $SNR$ in this spectrum is not high enough to clearly establish the presence or absence of \ion{C}{1}.

In the case of the Ly$\alpha$ feature, we do not see the core due to the strong geocoronal emission and significant resonant scattering in the circumstellar and interstellar media. Emission from $-800 < v < 800$ km s$^{-1}$ is entirely absorbed by the wind, but the emission from $800-2000$ km s$^{-1}$ is strong. Emission at these velocities indicates large amounts of scattering by the dense material \citep[$n \sim 10^{15}$ cm$^{-3}$][]{Zhu_outburst_FUOri_2020MNRAS} surrounding the accretion shock. 

The \ion{C}{2} doublet is highly wind-absorbed and only the red wings of the line profiles are visible. The same wind absorption is seen in GM Aur or DE Tau \citep{Xu_CIILinesWind_2021ApJ}. The absorption in the lines extends to $> -200$ km s$^{-1}$, like the $i = 35^\circ$ systems observed by \citet{Xu_CIILinesWind_2021ApJ}. 

The \ion{C}{4} doublet line profiles are less analogous to their CTTS counterparts \citep{Ardila_hotLinesInCTTSs_2013ApJS}. The \ion{C}{4} profiles in GM Aur and DE Tau are relatively symmetric, whereas in FU Ori the blue component of the doublet cannot be clearly identified. It is possible that there is some wind absorption of \ion{C}{4}, which would point to the wind in FU Ori being hotter than in quiescent YSOs. This has been proposed to be the case for FU Ori objects by \citet{Carvalho_HBC722_2024arXiv240520251C} in their analysis of the visible range wind emission lines of HBC 722. \citet{Ardila_hotLinesInCTTSs_2013ApJS} also note that the only object in their sample for which the \ion{C}{4} lines have P Cygni profile is DX Cha, a Herbig Ae star that may have a wind temperature greater than $10^5$ K. 

The He II 1640.4 \AA\ line is narrow and appears redshifted to 50 km s$^{-1}$ with a FWHM of $\sim 150$ km s$^{-1}$. The line may be wind-absorbed so that we are only seeing red-shifted emission. Since the \ion{C}{4} lines tend to form at higher temperatures than the He II line, this is expected if there is indeed wind absorption is seen in \ion{C}{4}. 

The H$_2$ emission in FU Ori is very weak. There are a few features that are almost as bright as the \ion{C}{2} and \ion{C}{4} emission, but many of the H$_2$ emission lines seen in CTTSs are not detected in the COS or STIS FUV spectra (see the line locations marked in Figure \ref{fig:COSSpec}). The lines are almost all blue-shifted to $-30$ km s$^{-1}$. The features are narrow, with typical FWHM values of 10 km s$^{-1}$, although their profiles vary from feature to feature. This variation may point to multiple origins for the H$_2$ emission in the system. It is possible that the emission is weaker than might be expected for such a rapidly accreting system due to strong absorption of Ly$\alpha$ in the wind, preventing photons from reaching the H$_2$ to excite emission. Furthermore, the wind temperature estimated for FU Ori objects by \citet{Carvalho_HBC722_2024arXiv240520251C} is sufficiently high that combined with the fact that many of the features are blue-shifted, a significant fraction of the H$_2$ we observe may be from the wind.

One feature that is present in many CTTSs and clearly absent in the FU Ori spectrum is the "bump" at 1600 \AA\  \citep{France_H2Bump_2017ApJ}. It is suggested by \citet{France_H2Bump_2017ApJ} that the bump is due to H$_2$O dissociation by Ly$\alpha$ in the inner 2 AU of the disk. If so, the lack of a bump here indicates that either the Ly$\alpha$ flux does not reach the disk surface due to strong self-absorption in the outflow, as discussed above, or that the bump emission is not bright enough to compare with the shock continuum emission.

\section{Previous Analysis of the FU Ori NUV Spectrum}\label{sec:NUVMeas}


While we present the first detection of the FUV continuum in FU Ori, previous studies have used the 2001 STIS/E230M spectrum to estimate the potential for boundary layer emission in the system \citep{Kravtsova_FUOriSTIS_2007AstL}. Here, we compare our findings with these results.

Our 16,000 K best-fit shock temperature is close to that reported in \citet{LopezMartinez_FUOriCII_2014MNRAS} based on modeling of the \ion{C}{2}] 2325 \AA\ emission line and nearby \ion{Fe}{2}] and \ion{Si}{2}] emission line ratios. Their analysis finds that the lines arise from a region with $T_e = 15,135$ K and $n_e = 1.78 \times 10^{10}$ cm$^{-3}$, assuming the emission is optically thin. Such a low $n_e$ is typical of the densities expected in the funnel flow of a CTTS with magnetospheric accretion \citep{Calvet_FunnelFlowStructure_1998ApJ}. 

However, \citet{LopezMartinez_FUOriCII_2014MNRAS} also report that the lines they analyzed in FU Ori are all blue-shifted to $v = -45$ km s$^{-1}$ and have full-width at half-maximum (FWHM) values of 221 km s$^{-1}$. The velocity offset matches the blue-shifted wind absorption found in the \ion{Mg}{2} 2796/2803 \AA\ doublet \citep{Kravtsova_FUOriSTIS_2007AstL}, while the FWHM is comparable with the broadening seen in the red-shifted emission of the NUV \ion{Mg}{2} doublet and the H$\alpha$ line in the visible \citep{herbig_high-resolution_2003}. 

These facts together indicate that both sets of emission lines may be formed in a strong outflow near the star. The absorption in \ion{Mg}{2} and emission in \ion{C}{2}] may trace the launch points of the outflow, whereas the broader (and red-shifted in \ion{Mg}{2}) emission traces scattering and turbulence in the wind at larger distances. The outflow may include material heated by the shock, which would explain the similar temperatures of the two components. 

In the low resolution G230L spectrum we detect the \ion{C}{2}] 2325 \AA, \ion{Si}{2}] 2350 \AA, and \ion{Mg}{2} 2795/2802 \AA\ lines, previously identified by \citet{Kravtsova_FUOriSTIS_2007AstL} at high resolution. The doublet reported at 2506 \AA\ is the second-brightest feature in the G230L spectrum. We identify the feature as likely being the 2507/2509 \ion{Fe}{2} doublet, which has been previously studied in the Weigelt blobs of $\eta$ Carinae \citep{Johansson_etaCar_FeII_1993PhST, Johansson_FeIILaserEtaCar_2004A&A}. The doublet is blue-shifted at the same velocity as the \ion{Si}{2}] and \ion{C}{2}] features, so it likely traces the same material. The appearance of this doublet in FU Ori is mysterious because it is not seen in the spectra of any other YSOs. Nevertheless the temperature of the outflowing material near the shock may be close to that of the Weigelt blobs \citep{Mehner_WeigeltBlobsDensityTemperature2010ApJ} and we expect the density is higher here than in the blobs.

\section{Revisiting the $\dot{M}$ and $A_V$ of FU Ori} \label{app:MdotAvTest}
In addition to modeling the excess FUV emission, we sought to also revisit the literature $\dot{M}$ and $A_V$ estimates for FU Ori. Using a recent SED model produced from a radiation-magneto-hydro-dynamical simulation of the system and an updated system inclination \citep{Perez_FUOriALMA_2020ApJ}, \citet{Zhu_outburst_FUOri_2020MNRAS} reported that the system has a $\dot{M} = 10^{-4.42} \ M_\odot$ yr$^{-1}$ and an $A_V = 1.5$ mag. Using $\dot{M} = 10^{-4.42} \ M_\odot$ yr$^{-1}$ in the disk $+$ shock model causes it to greatly overestimate the continuum redward of 2100 \AA, even in the visible range, due to the Rayleigh-Jeans tail of the shock emission. 

\begin{figure}[!htb]
    \centering
    \includegraphics[width = 0.5\linewidth]{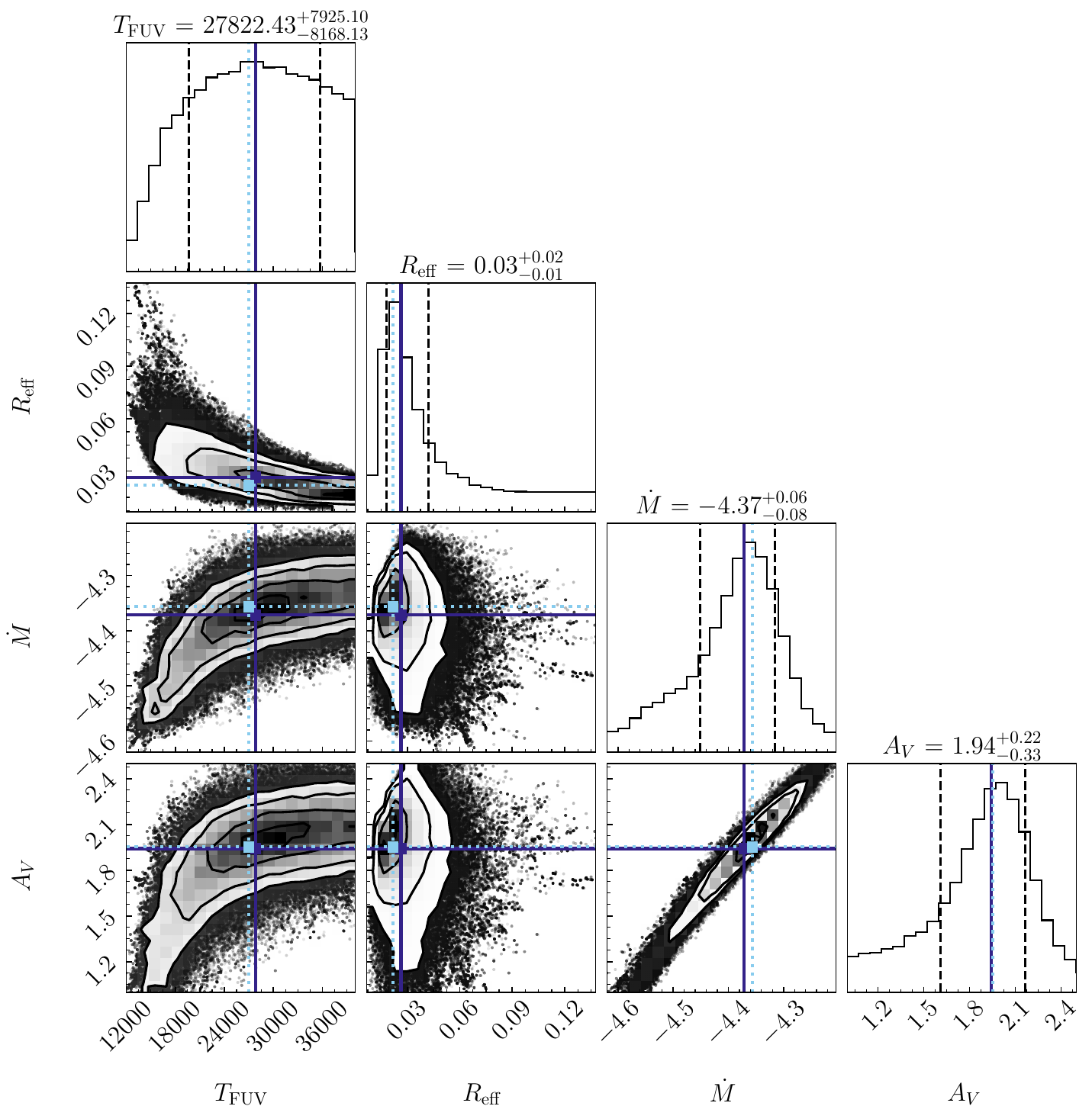}
    \caption{The $\mathtt{corner}$ plot for the fit with a variable $T_\mathrm{FUV}$, $R_\mathrm{eff}$, $\dot{M}$ and $A_V$. The dark blue vertical and horizontal solid lines mark the median values of the histograms, while the light blue dotted lines mark the modal values. The agreement between the two in all but the $T_\mathrm{FUV}$ parameter show that the posterior distributions are well-behaved. The contours in the 2-d histogram mark the 0.5, 1.0, 1.5, and 2.0 $\sigma$ levels, while the vertical dashed lines mark the 16th and 84th percentiles in the 1-d histograms.}
    \label{fig:corner-app}
\end{figure}

Using the same SED fitting procedure we describe in Section \ref{sec:cont_fit}, we allowed the $\dot{M}$ and $A_V$ to vary along with the $T_\mathrm{FUV}$ and $R_\mathrm{eff}$. We require that $T_\mathrm{FUV} < 20,000$ K in this fit due to the slight degeneracy between $T_\mathrm{FUV}$, $\dot{M}$, and $A_V$. When higher temperatures are allowed, the peak of the shock emission shifts blueward, requiring a higher $A_V$ to match the shape of the FUV continuum and a higher corresponding $\dot{M}$ to match the NUV. This produces a poorer overall fit to the UV and visible spectrum, especially in the 2100 to 2400 \AA\ and 3000 to 3800 \AA\ regions. The posterior distributions of the $\dot{M}$, $T_\mathrm{FUV}$, and $R_\mathrm{eff}$, are well-constrained, as shown in Figure \ref{fig:corner-app}. The $A_V$ posterior distribution is poorly constrained, although the fit clearly favors $A_V < 2$ mag. The $A_V = 1.5$ from previous literature, derived from fits to the visible range data \citep{zhu_FUOriInnerDisk_2007ApJ}, is well within the $1\sigma$ uncertainty of the best-fit $A_V$ here. Given the good agreement between our best-fit $\dot{M}$ and the existing literature value, we adopt $\dot{M} = 10^{-4.49} \ M_\odot$ yr$^{-1}$ for our fiducial model. Due to the poor constraint on $A_V$, we simply adopt the historical value of $A_V = 1.5$. 

\section{Detecting the Continuum Blueward of 1200 \AA} \label{app:COSCont}

In Section \ref{sec:discussion}, we stated that we observe emission blueward of Ly$\alpha$, which rules out the two photon process as the dominant source of FUV emission. In order to confirm the significance of the continuum detection in this range, we examined the background-subtracted count rate of the two G130M spectra ($\lambda_\mathrm{cen} = 1222$ \AA\ and $\lambda_\mathrm{cen} = 1309$ \AA). 

\begin{figure}
    \centering
    \includegraphics[width = \linewidth]{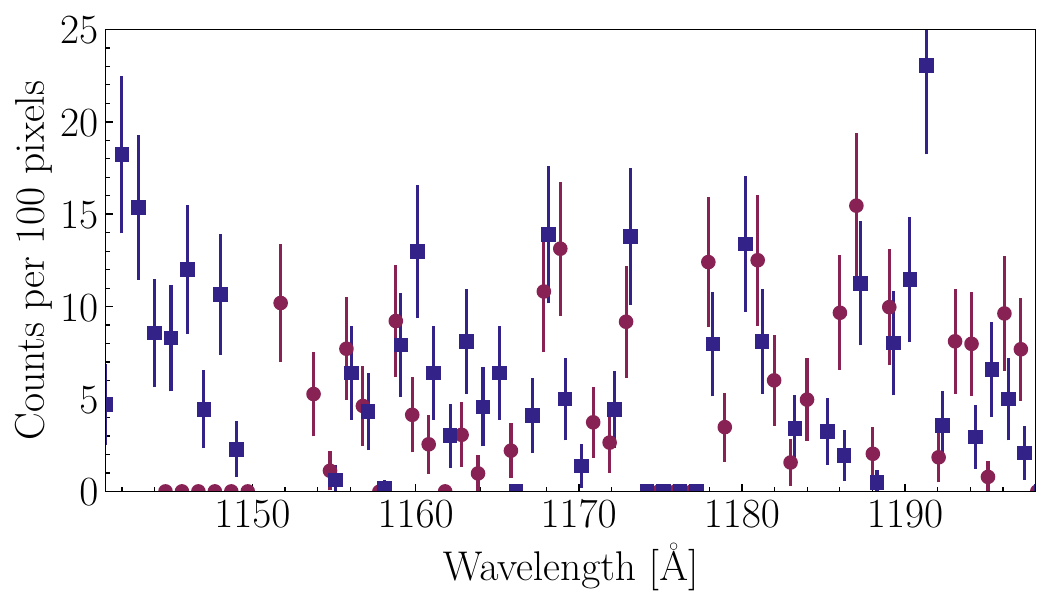}
    \caption{The net counts in the COS G130M spectra from 1150 \AA\ to 1200 \AA. The $\lambda_\mathrm{cen} = 1309$ \AA\ points are shown in maroon and the $\lambda_\mathrm{cen} = 1222$ \AA\ points are shown as dark blue squares. The counts have been resampled to a grid with 100-pixel-wide bins. Prior to binning, we masked the spectrum following the procedure described in Section \ref{sec:data}. We assume the uncertainties on each bin are Poisson-distributed. Notice that there are several bins with $\sim 10$ counts, indicating a $> 3 \ \sigma$ continuum emission detection.}
    \label{fig:counts}
\end{figure}

We compute the total net counts during the integration by multiplying the count rate by the 2348 s and 2350 s exposure times. We then mask the wavelengths that are known to have emission lines, following the procedure given in Section \ref{sec:data}. We then resample the data onto a grid with bin widths of 100 pixels each. We assume that the count rates are Poisson-distributed to estimate the uncertainties. We show the resulting spectrum in Figure \ref{fig:counts}.

The continuum emission at $\lambda < 1216$ \AA\ is detected in most of the bins with several exceeding 9 counts, or $3 \sigma$ assuming Poisson-distributed noise. If we bin all counts blueward of 1200 \AA\ into a one, the detection significance rises to $10 \sigma$. The two photon process has a sharp emission cut-off at $\sim 1215$ \AA, so if it cannot be the source of the continuum we see in this wavelength range. Therefore, we rule-out the possibility that the two photon process produces the observed FUV excess in FU Ori.

\end{document}